\documentclass[prl,twocolumn,unsortedaddress,showpacs]{revtex4}
\usepackage{graphicx}
\usepackage{hyperref}
\begin{document}
\title{Ion collection by oblique surfaces of an object\\ in a
  transversely-flowing strongly-magnetized plasma}
\author{I H Hutchinson}
\email{ihutch@mit.edu}
\homepage{http://www.psfc.mit.edu/~hutch/}
\affiliation{Plasma Science and Fusion Center and 
Department of Nuclear Science and Engineering,\\
 Massachusetts Institute of Technology, Cambridge, MA 02139, USA}

%Submitted to arxiv.org, 1 Jan 2008.
%Article-id: 0801.0309, Article password: w9iwk

\begin{abstract}
The equations governing a collisionless obliquely-flowing plasma around an
ion-absorbing object in a strong magnetic field are shown to have an
exact analytic solution even for arbitrary (two-dimensional) object-shape, when
temperature is uniform, and diffusive transport can be ignored. The
solution has an extremely simple geometric embodiment. It shows that
the ion collection flux density to a convex body's surface depends only upon
the orientation of the surface, and provides the theoretical
justification and calibration of oblique `Mach-probes'. The
exponential form of this exact solution helps explain the approximate
fit of this function to previous numerical solutions.
\end{abstract}
\maketitle

Ion collection by solid objects is key to a variety of phenomena
including the charging of dust\cite{shukla02} and
spacecraft\cite{hastings04}, and the measurement of plasma parameters
by electric probes\cite{HutchinsonBook2002}. In strongly magnetized
situations, where the ion Larmor radius is much smaller than the
object, the ion collection is highly anisotropic. The ion motion
parallel to the magnetic field can be approximated as governed by
free-flowing compressible hydrodynamic equations, while perpendicular
motion is slower, diffusive or, as in the present work, arising from
impressed transverse flow. For convenience the object is here
considered to be stationary and the plasma flowing, but this is
obviously equivalent to motion of the object in a stationary plasma. A
remarkable theoretical result is derived here consisting of a direct
analytic solution of the equations governing collisionless ion
collection when the plasma has fixed temperature, fixed transverse
velocity, negligible transverse diffusion, and arbitrary parallel
velocity in the external, uniform unperturbed region. This solution
helps resolve remaining questions concerning the calibration of
so-called Mach-probes, for measuring plasma flow, and provides
analytic expressions for the collection flux density as a function of
position on a (two-dimensional) object of essentially arbitrary shape.

When, as is typical, the Debye length is short compared with other
lengths in the problem, the interesting plasma region is quasineutral.
The charged thin sheath surrounding the object can be ignored except
in so far as it imposes a boundary condition upon the plasma
`presheath' solution: the `Bohm condition'\cite{riemann91}. If the ion Larmor
radius is also negligibly small, the `magnetized presheath' boundary
layer where ions are accelerated across the magnetic field can also be
ignored. The plasma transport anisotropy arising from the magnetic field causes
the presheath perturbation to the (uniform) background plasma to
elongate along the magnetic field until cross-field divergence makes
up the parallel flow to the object\cite{HutchinsonBook2002}. 
Full solutions to the non-linear
equations for \emph{diffusive} transverse transport have been obtained
numerically\cite{hutchinson87,Hutchinson1988}. 
The scaled solution, and hence the collected
flux, depends only upon the ratio of momentum diffusivity
to particle diffusivity. For given ratio (usually unity), it yields
the dependence of the collected ion flux density on the parallel
Mach-number, which amounts to the calibration factor for a (parallel)
Mach-probe, when the external perpendicular velocity is ignorable.

To measure the perpendicular plasma velocity, when it is not
negligible, one can use oblique collection faces, which show
sensitivity to that flow. The calibration of that sensitivity was
calculated\cite{VanGoubergen1999} using an extension of the
one-dimensional approximation of \cite{hutchinson87,Hutchinson1988},
still treating the perpendicular transport as diffusive, but
accounting for the boundary-condition
modification\cite{Hutchinson1996} that the transverse flow
causes. This solution can be shown to be a Galilean transformation of
the solution for zero transverse flow\cite{HutchinsonBook2002}. So it
is appropriate for an oblique surface of effectively infinite
dimension in the transverse flow direction, but finite in the
direction perpendicular to both flow and magnetic field. Practical
Mach-probes, in contrast, are more often multi-faceted `Gundestrup'
type\cite{MacLatchy1992,peleman06}, where many short adjacent
collectors are used with different orientations. So it is not obvious
theoretically that the diffusive solution should apply. And indeed the
opposite limit is probably more appropriate, where diffusion (and
viscosity) is ignored in comparison with the transverse flow. That
alternative is solved here.

The continuity and parallel momentum equations for the ion fluid in
steady state are
\begin{eqnarray}\label{governing}
M_\perp{\partial n \over \partial y} + 
 {\partial\over \partial x}\left(n M\right) =0 &,\nonumber\\
n M_\perp{\partial M \over \partial y} +
n M {\partial M \over \partial x} +
{\partial n \over \partial x} = 0 &,
\end{eqnarray}
where $M$ and $M_\perp$ are respectively the parallel (${x}$-) and
perpendicular (${y}$-) velocities normalized to the sound speed, $c_s$,
(i.e. the Mach numbers), and n is the density. The $z$-coordinate is
ignorable. The perpendicular velocity $M_\perp$ is taken as a constant
(positive without loss of generality) which is appropriate for uniform
external flow. This perpendicular velocity assumption is the only
essential difference between this problem and the \emph{unmagnetized}
quasineutral fluid-plasma problem, which can also be analysed analytically
for supersonic flow past a flat plate or
wedge\cite{lam66,wang92}. In the outer unperturbed region far from
the object, $n=n_\infty$ and $M=M_\infty$. The assumption of uniform
temperature is physically reasonable when the pressure term mostly
arises from the electrons, which are assumed Boltzmann-distributed.

These homogeneous equations can be placed into the following form
which displays their `characteristics':
\begin{eqnarray}\label{characteristics}
\left[M_\perp{\partial \over \partial y} + 
 (M+1){\partial\over \partial x}\right](\ln n +M) =0 &,\nonumber\\
\left[M_\perp{\partial \over \partial y} +
 (M-1){\partial\over \partial x}\right](\ln n -M) =0 &.
\end{eqnarray}
The solution thus requires both
\begin{eqnarray}
(\ln n +M)&=&const\mbox{\quad along }dx=dy (M+1)/M_\perp ,\label{positive}\\
\noalign{\noindent and}
(\ln n -M)&=&const\mbox{\quad along }dx=dy (M-1)/M_\perp .\label{negative} 
\end{eqnarray}
These will
be referred to respectively as the positive and negative
characteristics.

\begin{figure}[ht]
\medskip
\includegraphics[width=\hsize]{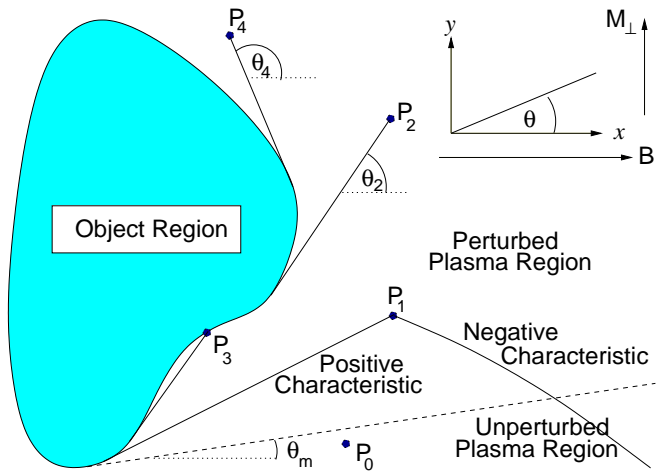}
\caption{\label{contsoln}Construction of the solutions at different
  points. $P_0$ is in the unperturbed region. For $P_1$ the
  characteristics are shown. $P_3$ is in a concave region and so its
  positive characteristic is not tangent at $P_3$. A value $\theta >
  \pi/2$ such as for $P_4$ is not problematic.
}
\end{figure}

For definiteness, we now consider plasma that is on the higher-$x$
side (to the right) of the object. Figure \ref{contsoln} will be used
for illustration. For any point in the plasma, a positive and a
negative characteristic pass through it. If we track the
characteristics backward and find that they both originate in the
unperturbed plasma at $y\to-\infty$, then we know that both the
conditions $\ln n + M=\ln n_\infty+M_\infty$ and $\ln n - M=\ln
n_\infty -M_\infty$ (where $_\infty$ indicates values in the
unperturbed region) are satisfied at that point. These simultaneous
equations have only one solution: $n=n_\infty$, $M=M_\infty$. The
point is in the unperturbed region if both its characteristics
originate there; for example: $P_0$. The characteristics for such
points are straight lines with slopes $M_\perp/(M_\infty\pm 1)$.

Next we consider the possibility that just one of the characteristics
originates not at $y=-\infty$, but on the surface of the object. We
notice that since the positive characteristic (\ref{positive}) always
has more positive $dz$ than the negative (\ref{negative}), the
positive characteristic is always to the left of the negative
characteristic, when tracing backward from a common point. Thus the
positive characteristic will always encounter the object first; so the
characteristic that originates on the object is the positive one. On
that characteristic, $(\ln n + M)$ is constant, but not equal to the
unperturbed value. However, because of the negative characteristics,
the parameters do satisfy $\ln n - M=\ln n_\infty -M_\infty$ both at
the point under consideration (e.g. $P_1$), and along the full length
of its positive characteristic. The only way to satisfy these two
requirements is that, along the positive characteristic, $M=const$ and
$n=const$. If $M=const$, then the slope of the characteristic,
$M_\perp/(M+1)$ is constant. It is a straight line.

The remaining question is, which straight-line is the positive
characteristic? This can be answered only by taking into account the
condition at the boundary between the plasma and the object. The
physics of the ion absorption, when the object is negatively charged,
is approximated by taking ions to flow into it as if they were flowing
into a vacuum. This requires the ions to flow into the surface as fast
as they can, consistent with existence of an overall solution. In
other words, at the object boundary, $M$ must be as negative as
possible consistent with the overall solution. The most negative $M$
is equivalent to the greatest slope-angle
$\theta\equiv\arctan[M_\perp/(M+1)]$ (even perhaps such that $\theta >
\pi/2$).  Thus for points close to the boundary, the positive
characteristic selected is the line with largest $\theta$ that
intersects the boundary (when tracked backwards). Such an extremum is
always \emph{tangential} to the object boundary where it intersects
it. Thus all positive characteristics that originate on the boundary
do so as tangents, and for any point in the perturbed plasma region
\emph{the positive characteristic is that straight line passing through the
point which has greatest $\theta$ and originates as a tangent on the
object}. Once that line is determined geometrically, its slope
determines $M$ and hence $n$ at all points along it. If the steepest
tangent angle is less than $ \theta_m = M_\perp/(M_\infty+1)$, then
the positive characteristic does not intersect the object, but has
slope $\theta=\theta_m$; and the point is in unperturbed plasma. This
specifies the entire solution for an arbitrary-shaped object:
\begin{equation}\label{curved}
  n=n_\infty \exp(M-M_\infty),\quad M=M_\perp \cot\theta -1.
\end{equation}

Because the argument concerning the boundary condition is subtle, and
to connect this derivation to prior more specific solutions, an
alternative approach giving the same solution is now
presented. Consider first an object which presents to the plasma a
single plane face at angle $\theta$. The beginning of this face is
taken as the origin. [We can suppose that the face which connects to it
in an edge at the origin is parallel to the $x$-axis, although that is
not important.] It can be shown, as a straightforward generalization of
\cite{hutchinsonreply88} , that this system (with an absorbing
boundary) is self-similar. The solution depends only on the ratio
$x/y$. Actually two self-similar forms satisfy the equations: either
the trivial uniform solution $M=const$, $n=const$; or the non-uniform
solution, $M=M_\perp x/y \mp 1$, $\ln n = \pm M + const$ (with the
upper sign the correct choice for our configuration). The full
solution with the appropriate boundary conditions is obtained by
matching these forms by a slope-discontinuity along the line
$x/y=x/y|_m= (M_\infty+1)/M_\perp$. Thus for this single-faced
problem, for $x/y \le x/y|_m$,
\begin{equation}\label{oblique}
  n=n_\infty \exp(M-M_\infty),\quad M=M_\perp x/y -1,
\end{equation}
while for  $x/y > x/y|_m$, $M=M_\infty$, $n=n_\infty$.

\begin{figure}[ht]
\medskip
\includegraphics[width=\hsize]{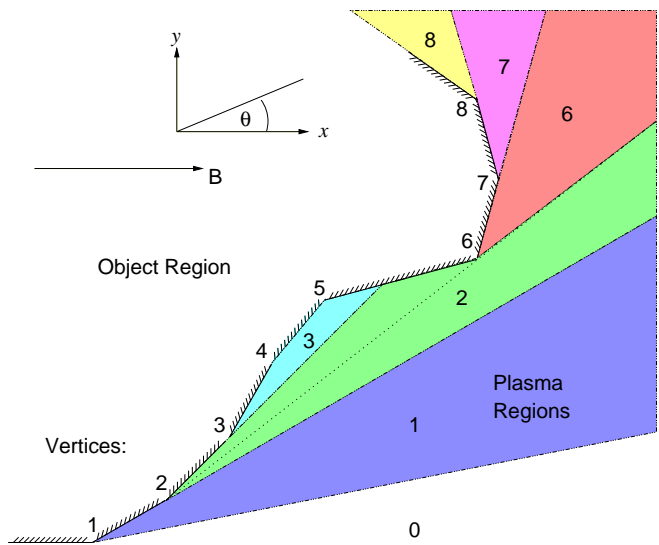}
\caption{\label{polygon}Geometric construction of the regions of applicability
  of the solutions for a multifaceted polygonal surface.}
\end{figure}

Now consider a multi-faceted object, as illustrated in figure
\ref{polygon}, where each (planar) facet and its leading (lower) vertex is
numbered. For the first facet (from vertices 1 to 2), the solution
(\ref{oblique}) just obtained applies, in region 1. The region 0 is
the unperturbed, uniform plasma, and the line separating regions 0 and
1 is the matching line $x/y= (M_\infty+1)/M_\perp$. Because the
$y$-velocity is fixed and positive, influence cannot propagate
downwards. So the solution on the first facet is unaffected by the
presence of different facets at higher $y$. At the trailing edge of
that facet (vertex 2) a transition to a different solution must occur.

Regard vertex 2 as the new origin, relative to which the coordinates
are $(x_2,y_2)=(x-p_2,y-q_2)$ (where $(p_i,q_i)$ are the coordinates
of vertex $i$). We can then, by just the same approach, construct a
self-similar solution to the governing equations (\ref{governing}) in
the region adjacent to facet 2; that solution depends only on the
ratio $x_2/y_2$. This new solution must be matched by a slope
discontinuity to the solution adjacent to the first facet. There is
just one line where that matching can be made, because on it
continuity of $M$ requires 
\begin{equation}\label{matchM}
  M_1=M_\perp x_1/y_1-1 = M_2=M_\perp x_2/y_2 -1,
\end{equation}
which shows $x_1/y_1=x_2/y_2$. The matching line is the
projection of facet 1 beyond its upper vertex. If facet 2 were of
lesser slope than facet 1, this match line would not be in the plasma
region; it would be inside the object. In that case there would in
fact be no second solution, the solution from facet 1 would run
directly into facet 2.

Clearly the process of matching solutions can be continued to
an arbitrary number of facets. The projections of the facets define
wedge-shaped regions in which the solution corresponding to each facet
terminates in a matching line. In region $i$, the complete solution is
\begin{equation}\label{solutioni}
 n=n_\infty \exp(M-M_\infty),\quad   M=M_\perp x_i/y_i -1,
\end{equation}
where $(x_i,y_i)\equiv(x-p_i,y-q_i)$. If a facet lies in a concave
region of the object, like facets 4 and 5 of figure \ref{polygon},
then it does not possess its own plasma region. Instead the solution(s)
of the earlier region(s) apply right up to the respective fractions of
that facet.

There is no difficulty with the angle of the facets exceeding
$\pi/2$. The solution (\ref{solutioni}) applies equally well in
regions 7 and 8 of figure \ref{polygon}, where $x_i/y_i$ is negative.
However, if the angle of the facets exceeds $\pi$, this solution does
not apply because then the plasma region would lie to the left of the
object, and the negative characteristic could not reach to the surface
from the right-hand side of the object. Normally, such facets are
simply part of the left-hand boundary of the object and are thus
governed by the appropriate left-hand version of eq (\ref{solutioni})
tracked round the left side of the object. If (part of a) facet cannot
be reached from either side of the object without passing through an
angle of $\pi$, then it corresponds to a region that has no plasma
flowing to it either from right or left. The density there is zero. No
characteristics originating in plasma pass through it.  Finally,
plasma in regions such that $y$ exceeds all points on the object are
said to be in the `wake'. Normally the wake begins on the upper-most
(rear-most) point of the object. At that point the solutions from the
right and the left collide, forming a shock, and the simplified
equations we are considering cannot be satisfied. For example, the
temperature cannot be the same in the shocked region. Since the wake
does not affect the ion collection by the object (in so far as
equations (\ref{governing}) apply), it is not further analysed here.

Obviously, the continuous curved-boundary solution (\ref{curved}) is
the straightforward limit of the multifaceted solution
(\ref{solutioni}) as the number of facets tends to infinity. The
discussion about facet angles, etc., applies directly to tangents for the
continuous case.

The solution (\ref{curved}) provides an extremely simple formula
for the ion flux to a surface not affected by concavity. Adding
the perpendicular and parallel components, the total flux density is 
$n c_s(M_\perp \cos\theta - M \sin\theta) = n c_s\sin\theta$. Written
as flux per unit area perpendicular to the magnetic field, this is
\begin{equation}\label{flux}
  \Gamma_\parallel = n_\infty c_s \exp[-1 -(M_\infty- M_\perp\cot\theta)].
\end{equation}

This form indicates, first, that for points not in a concave region of
the object, the collected flux depends only on the angle of the
surface there, and not on the shape of the object at smaller $y$. This
is an important, but not obvious, result for establishing Mach-probe
performance.  Second, the exponential dependence upon $M_\infty$ is
within 10\% of the dependence, $\exp[-1-1.1(M_\infty -
  M_\perp\cot\theta)]$, that fits the diffusive
solution\cite{VanGoubergen1999,HutchinsonBook2002}; the only
difference being to replace 1.1 with 1. This agreement, well within
most experiments' uncertainties, is an
indication that the calibration is not strongly dependent upon the
assumed cross-field transport regime. Moreover, the present solution
helps to explain why the exponential form is such a good fit to the
prior numerical solutions, and to give confidence that it is
reasonable to extend the form to supersonic velocities, which in the
present context are unproblematic. Furthermore the present solution
shows unambiguously that leading faces, for which $\theta<\theta_m$
receive simply the unperturbed flux [$\Gamma_\parallel = n_\infty c_s
  (M_\perp \cot\theta - M_\infty)$], while trailing faces, even those
for which $M_\infty- M_\perp\cot\theta>1$, are governed by the formula
(\ref{flux}). The boundary conditions at the magnetic presheath edge
arise here naturally from the analysis of the quasi-neutral
equations. No prior assumption about the boundary velocity is required,
although the condition is, of course, consistent with prior
treatments.

In a concave region of the object, e.g. at $P_3$, the surface angle (local tangent)
$\theta_s$, is smaller than the characteristic's angle $\theta$, and the
distinction must be retained. This leads to an enhanced ion flux, equal to
equation (\ref{flux}) times the extra factor
$(M_\perp\cot\theta_s-M_\perp\cot\theta +1)$.

For the equations used, the present solutions are exact. The
formulation is, of course, approximate. Modifications both to the
boundary conditions and to the equations may be appropriate in some
practical applications. Nevertheless, the present analysis captures
the dominant physics of the situation and helps to clarify both the
operation of oblique Mach-probes and the ion collection by objects of
essentially arbitrary shape when the ion Larmor radius is negligible.

\bibliography{hutch}

\end{document}